\documentclass[a4paper,11pt]{article}

\usepackage{jheppub} 
\usepackage[T1]{fontenc}
\newcommand{\be}{\begin{equation}}
\newcommand{\ee}{\end{equation}}

\newcommand{\bea}{\begin{eqnarray}}
\newcommand{\eea}{\end{eqnarray}}
\newcommand{\bfig}{\begin{figure}}
\newcommand{\efig}{\end{figure}}
\newcommand{\bc}{\begin{center}}
\newcommand{\ec}{\end{center}}

\DeclareFontFamily{U}{wncy}{}
\DeclareFontShape{U}{wncy}{m}{n}{<->wncyr10}{}
\DeclareSymbolFont{mcy}{U}{wncy}{m}{n}
\DeclareMathSymbol{\sha}{\mathord}{mcy}{"58}

\title{Two-Loop integrals for CP-even heavy quarkonium production and decays: Elliptic Sectors}

\author[a]{Long-Bin Chen,}
\author[b]{Jun Jiang,}
\author[b,c,1]{Cong-Feng Qiao\note{Corresponding author.}}

\affiliation[a]{School of Physics \& Electronic Engineering, Guangzhou University, Guangzhou 510006, China}
\affiliation[b]{School of Physics, University of Chinese Academy of Sciences,\\YuQuan Road 19A, Beijing 100049, China}
\affiliation[c]{CAS Center for Excellence in Particle Physics, Beijing 100049, China}

\emailAdd{chenglogbin10@mails.ucas.ac.cn}
\emailAdd{jiangjun13b@mails.ucas.ac.cn}
\emailAdd{qiaocf@ucas.ac.cn}

\abstract{By employing the differential equations, we compute analytically the elliptic sectors of two-loop master integrals appearing in the NNLO QCD corrections to CP-even heavy quarkonium exclusive production and decays, which turns out to be the last and toughest part in the relevant calculation. The integrals are found can be expressed as Goncharov polylogarithms and iterative integrals over elliptic functions. The master integrals may be applied to some other NNLO QCD calculations about heavy quarkonium exclusive production, like $\gamma^*\gamma\rightarrow Q\bar{Q}$, $e^+e^-\rightarrow \gamma+ Q\bar{Q}$,~and~$H/Z^0\rightarrow \gamma+ Q\bar{Q}$, heavy quarkonium exclusive decays, and also the CP-even heavy quarkonium inclusive production and decays.}

\keywords{QCD, Quarkonium, Loop integrals, Polylogarithms, Elliptic integrals}
\preprint{~}

\begin{document}
\maketitle
\flushbottom

\section{Introduction}

Precision physics in colliders requires more higher-order corrections in perturbation theory. Unravelling the mathematical structure of Feynman integrals in multiloop calculation is somehow critical to handle the complexity of higher order calculations and may help us to obtain a better control of the perturbative expansion. In recent years, the corresponding research achieved some breakthroughs and becomes now one of the hot topics in physics and mathematics.

One of the powerful methods to evaluate the master integrals analytically attributes to the differential equation \cite{Kotikov:1990kg, Kotikov:1991pm, Remiddi:1997ny, Gehrmann:1999as, Argeri:2007up}. With recent developments \cite{Henn:2013pwa,Henn:2013nsa,Henn:2014qga,Argeri:2014qva,Liu:2017jxz}, this method becomes now a prevailing one in tackling those integrals unsolvable before. It was noticed by Henn that generically in multi-loop calculation, choosing a set of suitable basis for master integrals can greatly simplify the corresponding differential equations \cite{Henn:2013pwa}, which can be calculated iteratively in dimensional regularization scheme. In light of this proposal, many of multi-loop Feynman integrals for various phenomenological processes have been calculated \cite{Henn:2013woa,Henn:2014lfa,Gehrmann:2014bfa,Caola:2014lpa,DiVita:2014pza,Bell:2014zya,Huber:2015bva,Chen:2015csa, Bonciani:2015eua,Gehrmann:2015dua,Grozin:2015kna,Bonciani:2016ypc,Becchetti:2017abb}.
Note, some Feynman integrals in two-loop or higher order possess new mathematical structures \cite{Ablinger:2017bjx,Ablinger:2014bra,Bloch:2013tra,Bloch:2014qca,Broedel:2017kkb,Broedel:2017siw,Broedel:2018iwv,Hidding:2017jkk}, which cannot be expressed as multiple polylogarithms and ask for different technique to deal with. A typical example is the massive two-loop sunrise integral, which has been studied intensively \cite{Laporta:2004rb,Kniehl:2005bc,Adams:2015ydq,Adams:2014vja,Adams:2015gva,Remiddi:2013joa,Adams:2016xah,Remiddi:2016gno,adams:2018}.

The heavy quarkonium production and decay are one of the hot topics in particle physics ever since the first discovery in 1974, especially with the advent of Nonrelativistic Quantum Chromodynamics (NRQCD) factorization formalism \cite{NRQCD}. Up to date there still exist some discrepancies between experimental data and theoretical expectations  \cite{Abe:2002rb,Aubert:2005tj,Zhang:2005cha,Zhang:2006ay}, which appeal for precision calculations. In one of our previous works \cite{Chen:2017xqd} we gave out a set of 86 two-loop  master integrals about heavy quarkonium production and decay, which can be cast into the canonical form and expressed in terms of multiple polylogarithms. However, for those Feynman integrals with functions beyond the realm of multiple polylogarithms the calculation is not done yet. In fact, to date, only a limited number of similar calculations have been performed in the literature.

In this work, we calculate analytically all remaining integrals with different mathematical structures from multiple polylogarithms in CP-even heavy quarkonium production and decays. The master integrals will be classified into two sectors, one with integrals containing sub-topologies related to the two-loop massive sunrise integrals and the other involving non-planar two-loop three-point integrals. Following the strategy suggested in Ref. \cite{Remiddi:2016gno} and with properly chosen basis, we cast the differential equations of those integrals in the first sector into a proper form that can be solved recursively. Of the second sector, the key point is to find the homogeneous solutions for the second-order differential equations of the two-loop non-planar three-point massive integrals, with that the full solutions can then be obtained by constant variation.

The paper is organized as follows. In section 2, the kinematics is discussed and the derivatives with respect to kinematic variables will be given. In section 3, the iterative integrals and complete elliptic integrals are introduced. In section 4, the elliptic type integrals will be separated into two sectors, and the calculation procedure for them will be elucidated respectively. For illustration, specific examples will be given. Section 5 is remained for conclusions and outlooks. The definition of master integrals is given in appendix A, and several simple but typical analytical results are presented in appendix B.

\section{Notation and kinematics}
\begin{figure}[t]
\begin{center}
\includegraphics[scale=0.5]{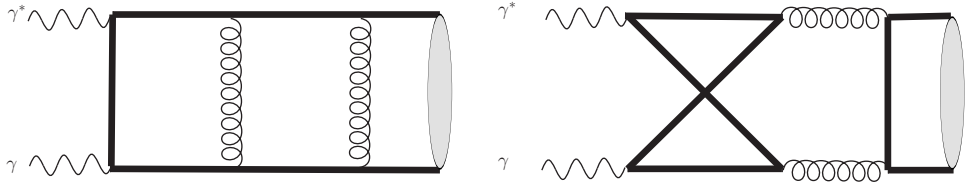}
\caption{Typical two-loop Feynman diagrams for CP-even heavy quarkonium production.}
\label{sample}
\end{center}
\end{figure}

The heavy quarkonium exclusive production in electron-positron collision has a relatively low background, and has played an important role in the study of quarkonium production mechanism. Here we calculate the CP-even quarkonium production in two correlated processes, that is in $\gamma^*\gamma$ collision and in electron-position annihilation associated with a photon,
\bea
\gamma^*(k_1)+ \gamma (k_2) \rightarrow  Q(k_q) \bar{Q}(k_{\bar{q}})\ ,\label{pro1} \\
\gamma^*(k_1)\rightarrow Q(k_q) \bar{Q}(k_{\bar{q}}) + \gamma (k_2)\ ,\label{pro2}
\eea
where $k_1^2 = 2 ss , k_2^2 = 0$ and $k_q^2 = k_{\bar{q}}^2 = m_q^2$. The typical Feynman diagrams are showed in Fig. \ref{sample}. The process (\ref{pro1}) is in Euclidean region with $ss<0$, and the momenta satisfy the following relations
\be
(k_1+ k_2)^2 = (k_q + k_{\bar{q}})^2 = 4 m_q^2\ .
\ee
Whereas, the process (\ref{pro2}) is in Minkowski region  with $2ss > 4 m_q^2$,  and
\be
(k_1- k_2)^2 = (k_q + k_{\bar{q}})^2 = 4 m_q^2\ .
\ee
Note, in the threshold expansion approach, quark and anti-quark momenta are taken to be equal, i.e. $k_q = k_{\bar{q}}$.

In order to express the results compactly, here we introduce three dimensionless variables $x$, $y$ and $z$ as follows:
\bea
\frac{ss}{m_q^2} = -\frac{(1-x)^2 }{2x}=(y+2)=(z+1) \ .\label{xyz}
\eea

The NNLO QCD corrections to processes (\ref{pro1}) and (\ref{pro2}) are calculated in light of Feynman diagrams. As a routine, with some algebraic manipulations, the amplitudes can be reduced to a set of scalar integrals. We use the Mathematica package $\textbf{FIRE}$ \cite{Smirnov:2008iw,Smirnov:2013dia,Smirnov:2014hma} to reduce the scalar integrals to a minimum set of independent master integrals. The calculation of these master integrals is the central issue, and normally turns out to be a nontrivial work. In our calculation, we apply the method of differential equations to calculate the master integrals.

The first step of deriving differential equations is taking derivatives of the Lorentz invariant kinematic variables, and expressing them as linear combinations of master integrals. The $\textbf{FIRE}$ is also employed in the derivation of
differential equations. The derivatives of the external momenta can be expressed as the derivatives of $ss$ and $m_q^2$, like
\bea
k_i\cdot\frac{\partial}{\partial k_j}=k_i\cdot\frac{\partial ss}{\partial k_j}\frac{\partial}{\partial ss} + k_i\cdot\frac{\partial m_q^2}{\partial k_j}\frac{\partial}{\partial m_q^2}
\eea
with $i(j)=1\ \text{or}\ 2$. And in reverse, the derivative $\frac{\partial }{\partial ss}$ can be expressed as a linear combination of derivatives $k_i\cdot\frac{\partial}{\partial k_j}$, i.e.,
\bea
2 ss \frac{\partial}{\partial ss} = k_1\cdot \frac{\partial}{\partial k_1} + \left(\frac{ss + 2 m_q^2}{ss - 2 m_q^2}\right) k_2\cdot \frac{\partial}{\partial k_2}\ .
\eea
The derivative transform can be readily obtained according to equation (\ref{xyz}). With the variables chosen in above,
analytical results of the integrals can then be formulated in a compact form, in terms of iterative integrals and elliptic integrals.

\section{Iterated integrals and complete elliptic integrals}

The Goncharov polylogarithms (GPLs) \cite{Goncharov:1998kja} are defined as
\bea
G_{a_1,a_2,\ldots,a_n}(x) &\equiv & \int_0^x \frac{\text{d} t}{t - a_1} G_{a_2,\ldots,a_n}(x)\ ,\\
G_{\overrightarrow{0}_n}(x) & \equiv & \frac{1}{n!}\log^n x\ ,
\eea
which in fact are special cases of a more general type of integrals, named Chen-iterated integrals \cite{Chen}. If all indices $a_i$ belong to set $\{0, \pm 1\}$, the Goncharov polylogarithms can then be transformed into the well-known Harmonic polylogarithms (HPLs) \cite{Remiddi:1999ew}
\bea
H_{\overrightarrow{0}_n}(x) &=&G_{\overrightarrow{0}_n}(x)\ ,\\
H_{a_1,a_2,\ldots,a_n}(x) &=&(-1)^k G_{a_1,a_2,\ldots,a_n}(x)\ ,
\eea
where $k$ equals to the number of times the element $(+1)$ appearing in $(a_1,a_2,\ldots,a_n)$\, .
The GPLs satisfy the following shuffle rules:
\bea
G_{a_1,\ldots,a_m}(x)G_{b_1,\ldots,b_n}(x) &=& \sum_{c\in a \sha b} G_{c_1, c_2,\ldots,c_{m+n}}(x)\ .
\eea
In above equation, $a \sha b$ is composed of the shuffle products of $a_i (i=1,2\ldots m)$ and $b_i (i=1,2\ldots n)$, which is defined as the set of lists containing all elements of $a_i$ and $b_i$, with the order of elements
$a_i$ and $b_i$ preserved. The GPLs and HPLs can be numerically evaluated by implementing the {\bf GINAC}  \cite{Vollinga:2004sn,Bauer:2000cp}, and the Mathematica package {\bf HPL} \cite{Maitre:2005uu,Maitre:2007kp} is applicable
to the HPLs reduction and evaluation. Both GPLs and HPLs can be transformed into functions $\ln,\ \text{Li}_n$ and $\text{Li}_{22}$ up to weight four in light of the method described in Ref. \cite{Frellesvig:2016ske}.

In our calculation, the complete elliptic integrals are necessary to express the integrals encountered. The first and second kinds of complete elliptic integrals are defined as
\bea
K(x)=\int_0^1\frac{\text{d} t}{\sqrt{(1-t^2)(1-x~t^2)}}
\eea
and
\bea
E(x)=\int_0^1\frac{\sqrt{1-x~t^2}}{\sqrt{1-t^2}}\text{d} t\ .
\eea
They satisfy the following derivative relations:
\bea
\frac{\text{d} K(x)}{\text{d} x}&=&\frac{E(x)-(1-x)K(x)}{2(1-x)x}\ ,\nonumber\\
\frac{\text{d} E(x)}{\text{d} x}&=&\frac{E(x)-K(x)}{2x}\ .
\eea

The Legendre relation is useful in simplifying the complete elliptic integrals, i.e.,
\bea
K(x)K(1-x)-K(x)E(1-x)-E(x)K(1-x)=-\frac{\pi}{2}\ .
\eea

\section{Elliptic integral sectors }

The symbols and canonical basis in the calculation of elliptic integrals keep the same as in the preceding work \cite{Chen:2017xqd}, where the linear differential equations can be expressed, via a suitable basis choice of master
integrals, as canonical form \cite{Henn:2013pwa}
\bea
\text{d}~{\bf F} = \epsilon \, (\text{d}~{\bf A}) \, {\bf F} \,
\label{sysdeq}
\eea
with ${\bf F}$ being the vector of canonical master integrals $F_i (i=1\ldots 86)$ \cite{Chen:2017xqd}. Whereas, the two-loop massive Feynman integrals concerned in this work may involve elliptic functions, and hence the calculation of the integrals
should be further explored. We separate them into two elliptic sectors: one with integrals containing sub-topologies related to the two-loop massive sunrise integrals, the other with two-loop non-planar three-point integrals. In the following we elucidate the calculation procedures of these integrals.

\subsection{Sector I : integrals with massive sunrise integrals as subtopology}

\begin{figure}[t]
\begin{center}
\includegraphics[scale=0.46]{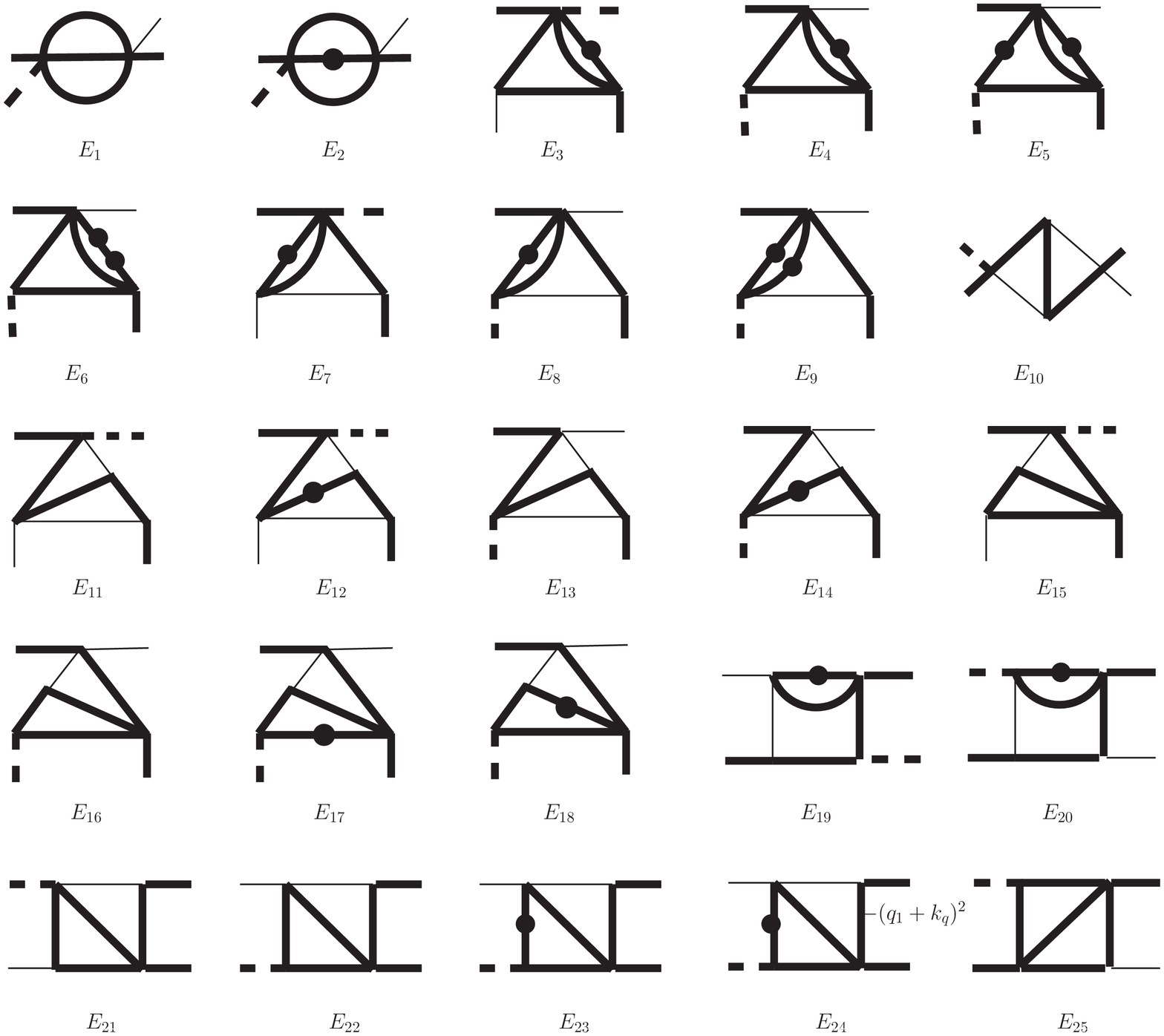}
\includegraphics[scale=0.46]{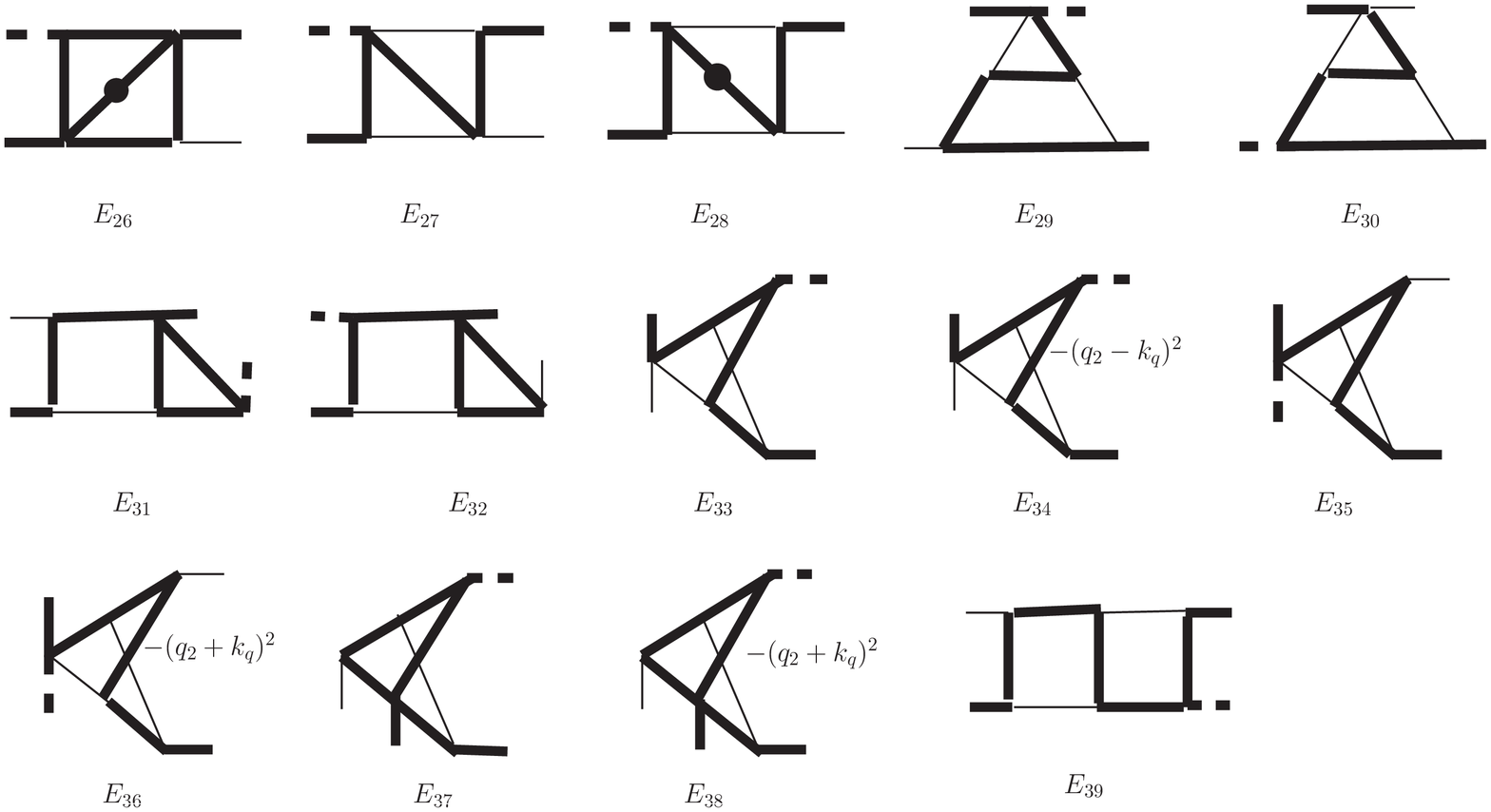}
\caption{The set of 39 master integrals involve elliptic functions in sector I. The thin line denotes massless propagators and on-shell massless external particles; the thick line represents massive propagators and on-shell massive external particles; the dash line indicates off-shell external particles with momentum squared equal to $2 ss$. The internal lines with a dot mean the power of the propagators are raised to 2.}
\label{midiag1}
\end{center}
\end{figure}

The $39$ Feynman integrals $E_i (i=1\ldots 39)$ belonging to this subsection are shown in Fig. \ref{midiag1}, which contain sub-topologies related to the two-loop massive sunrise integrals. The expressions of master integrals without numerators can be readily read off from the figure, and those with numerators are given in appendix A. Note, the massive sunrise integrals are composed of the complete elliptic integrals and cannot be expressed as pure Goncharov polylogarithms. The two-loop massive sunrise integrals ($E_1,E_2$) have been widely studied. Here, the bases $(\text{A}_1, \text{A}_2)$, which contain $(E_1, E_2)$, are of the same as their first appearance in Ref. \cite{Remiddi:2016gno}:
\bea
\text{A}_1 & = & \, \epsilon^2 \,\frac{12m_q^2((1-2\epsilon)(2(2-3\epsilon)E_1+2(ss+2m_q^2)E_2)-(ss-4m_q^2)F_1/\epsilon^2)}{(ss-2m_q^2)(ss-10m_q^2)} \, ,
\eea
\bea
\text{A}_2 & = & \, \epsilon^2 \, \frac{1}{m_q^2(ss-2m_q^2)(ss-10m_q^2)}\left(-8(1-2\epsilon)(2-3\epsilon)((1-4\epsilon)ss^2+4(11\epsilon-4)ss\, m_q^2 \right. \nonumber\\
& &  \left. + 4(3-10\epsilon)m_q^4)E_1 -8(1-2\epsilon)((2\epsilon-1)ss^3-6(7\epsilon-2)ss^2\, m_q^2+12(20\epsilon-7)ss\, m_q^4\right. \nonumber\\
& &  \left. -8(25\epsilon-8)m_q^6 )E_2-4((1-4\epsilon)ss^3+2(22\epsilon-3)ss^2 m_q^2\right.\nonumber\\
& &  \left. -4(3+10\epsilon)ss\, m_q^4+8m_q^6)F_1/\epsilon^2    \right)\ .
\label{a1a2}
\eea

In the following we sketch the calculation of this sector. With a suitable choice of the basis in the high topologies $(E_3\ldots E_{39})$, the homogeneous part of the differential equations for integrals $(E_3\ldots E_{39})$ can be cast into the canonical form, whereas depending on the inhomogeneous terms of massive sunrise integrals $(E_1,E_2)$, or $(\text{A}_1, \text{A}_2)$. To be more specific, after a proper selection of bases $\text{A}'_i(i=3\ldots39)$, the differential equations for $\text{A}'_i(i=3\ldots 39)$ can be expressed as
\bea
\frac{\text{d}\, \bf{A}'}{\text{d}\, ss} = \epsilon (\bf{W} \cdot \bf{A}' + \bf{Y} \cdot \bf{F}) + (\epsilon \bf{Q}_1 + \bf{Q}_2)  \text{A}_1+ \bf{Q}_3 \text{A}_2\ .
\label{de1}
\eea
Here, $\bf{A}'$ is a 37-dimensional basis vector containing integrals $E_i(i=3 \ldots 39)$ and $F_i(i=1 \ldots 86)$; $\bf{F}$ is a 86-dimensional basis vector that was given in Ref. \cite{Chen:2017xqd}; $\bf{W}$ and $\bf{Y}$ are $37\times37$ and  $37\times86$ matrices, respectively; $\text{A}_1$ and $\text{A}_2$ are scalar functions defined in equation (\ref{a1a2}); and $\bf{Q}_i$(i = 1,2,3) represent the 37-dimensional vectors which are composed of algebraic functions and are $\epsilon$ free.

Notice that in equation (\ref{de1}) the inhomogeneous term that contain $\text{A}_2$ is free of $\epsilon$, and the differential equation for $\text{A}_1$ given in Ref. \cite{Remiddi:2016gno} can be reexpressed as
\bea
\frac{\text{d}\, \text{A}_1}{\text{d}\, ss}&=&\frac{-(ss-m_q)^2+14(ss-m_q^2)m_q^2+3m_q^4}{2(ss-m_q^2)(ss-2m_q^2)(ss-10m_q^2)} \text{A}_1-\frac{2\epsilon}{ss-10m_q^2} \text{A}_1\nonumber\\
& &-\frac{3m_q^4}{2(ss-m_q^2)(ss-2m_q^2)(ss-10m_q^2)}\text{A}_2\ .
\eea
Since in above equation the inhomogeneous term containing $\text{A}_2$ is also $\epsilon$ free, we therefore are legitimate to perform a basis shift as
\bea
\text{A}'_i\rightarrow \text{A}'_i + b_i(ss)\text{A}_1 \equiv \text{A}_i\, (i=3\ldots 39)\ .
\eea
With the basis shift, $\bf{Q}_3 \text{A}_2$ will be removed from the differential equation (\ref{de1}). Here $b_i(ss)$ are algebraic functions to be determined. Moreover, the basis shift may also simplify the inhomogeneous term containing $\text{A}_1$, considerably.

For illustration, we take the differential equations for $(E_4,E_5,E_6)$ as an example, which have the same topology. By properly choosing the basis, the differential equations for $(E_4,E_5,E_6)$ can be formulated as
\bea
\frac{\text{d}\, e(ss,\epsilon)}{\text{d}\, ss} &=& \epsilon[Y^1(ss)e(ss,\epsilon)+W^1(ss)f(ss,\epsilon)]\nonumber\\
&&+\Omega^0(ss) \text{A}_{1}+\epsilon \Omega^1(ss) \text{A}_{1}+\Lambda^0(ss) \text{A}_{2}\ .
\eea
Here, $e(ss,\epsilon)$, a 3-dimensional basis vector containing integrals $(E_4, E_5, E_6)$ and $F_{12}$, may be expressed as
\begin{equation}
e(ss,\epsilon)=
\left(\begin{array}{c}e_1(ss,\epsilon)\\
e_2(ss,\epsilon)\\
e_3(ss,\epsilon)\end{array} \right)
=\left(
\begin{array}{c}
\epsilon^3 (ss-2m_q^2)  E_4\\
\epsilon^2 \sqrt{ ss}\sqrt{ss - 2 m_q^2}(ss-2m_q^2)E_5\\
\epsilon^2 \frac{ss-2m_q^2}{ss+2m_q^2}\left(6\epsilon m_q^2 E_4 - 4ss^2 E_5+m_q^2(3ss+2m_q^2)E_6\right)+\frac{6m_q^2}{ss+2m_q^2}F_{12}\\
\end{array}
\right),
\end{equation}
with $f(ss,\epsilon)$ being a 2-dimensional basis vector
\begin{equation}
f(ss,\epsilon)=\left(
\begin{array}{c}
F_8\\
F_{12}\\
\end{array}
\right)\ .
\end{equation}
$Y^1$ is a $3\times3$ matrix, $W^1$ is a $3\times 2$ matrix, $\{\Omega^0(x),\ \Omega^1(x),\ \Lambda^0(x)\}$ are 3-dimensional vectors, and $\text{A}_1$ and $\text{A}_2$ are scalar functions defined as (\ref{a1a2}).
To remove the $\text{A}_2$ dependence from the inhomogeneous part of the differential equations, we perform the basis shift
\bea
e_i(ss,\epsilon)\rightarrow e_i(ss,\epsilon)+b_i(ss) \text{A}_1\, (i=1,2,3)\ ,
\label{shift}
\eea
where $b_i(ss)$ are algebraic functions to be determined. By virtue of the differential equation for $\text{A}_1$,
one can figure out the shift functions $b_i(ss)$ in (\ref{shift}), which may be formulated in a 3-dimensional vector form
\begin{equation}
b(ss)=\left(
\begin{array}{c}
0\\
2\frac{(2ss-5m_q^2)\sqrt{ss - 2 m_q^2}}{3m_q^2\sqrt{ss}}\\
\frac{14ss m_q^2-13ss^2+8m_q^4}{3m_q^2(ss+2m_q^2)}\\
\end{array}
\right)\ .
\end{equation}
The differential equation for $e_1(ss,\epsilon)$ is in canonical form, and hence no need to make the shift. After the basis shift, $\Lambda^0(ss) \text{A}_{2}$  and $\Omega^0(ss) \text{A}_{1}$ terms in differential equation for $e_3(ss,\epsilon)$ vanish, and the differential equation for $e_3(ss,\epsilon)$ turns to be canonical. Of the differential equation for $e_2(ss,\epsilon)$, though $\Lambda^0(ss) \text{A}_{2}$ term does not exist, $\Omega^0(ss) \text{A}_{1}$ term remains. Note, with the basis shift the inhomogeneous part of the differential equations for $e_2(ss,\epsilon)$ will be greatly simplified, and the differential equations turn to be solvable recursively.

The method described above is also applicable to high sectors with more propagators. Except for integrals $(E_1,E_2, E_5, E_9 )$, differential equations for the remaining 35 integrals can be transformed into the canonical form (\ref{sysdeq}), with the method employed in this work. The basis vector ${\bf A}$ is built up with 39 functions $\text{A}_i(ss,m_q,\epsilon)$, the linear combinations of master integrals $E_i$ and $F_i$ with the latter given in Ref. \cite{Chen:2017xqd}. Explicitly, the 39 bases that contain planar and non-planar two-loop integrals can be formulated as
\bea
\text{A}_1 & = & \, \epsilon^2 \,\frac{12m_q^2((1-2\epsilon)(2(2-3\epsilon)E_1+2(ss+2m_q^2)E_2)-(ss-4m_q^2)F_1/\epsilon^2)}{(ss-2m_q^2)(ss-10m_q^2)} \, , \nonumber\\
\text{A}_2 & = & \, \epsilon^2 \, \frac{1}{m_q^2(ss-2m_q^2)(ss-10m_q^2)}\left(-8(1-2\epsilon)(2-3\epsilon)((1-4\epsilon)ss^2+4(11\epsilon-4)ss\, m_q^2 \right. \nonumber\\
& &  \left. + 4(3-10\epsilon)m_q^4)E_1 -8(1-2\epsilon)((2\epsilon-1)ss^3-6(7\epsilon-2)ss^2\, m_q^2+12(20\epsilon-7)ss\, m_q^4\right. \nonumber\\
& &  \left. -8(25\epsilon-8)m_q^6 )E_2-4((1-4\epsilon)ss^3+2(22\epsilon-3)ss^2 m_q^2\right.\nonumber\\
& &  \left. -4(3+10\epsilon)ss\, m_q^4+8m_q^6)F_1/\epsilon^2    \right)\ , \nonumber\\
\text{A}_3 & = & \, \epsilon^3 \,  (ss-2m_q^2)\,  E_3\ , \nonumber\\
\text{A}_4 & = & \, \epsilon^3 \,  (ss-2m_q^2)\,  E_4 \ , \nonumber\\
\text{A}_5 & = & \, \epsilon^2 \, \sqrt{ ss}\sqrt{ss - 2 m_q^2}(ss-2m_q^2)E_5+2\frac{(2ss-5m_q^2)\sqrt{ss - 2 m_q^2}}{3m_q^2\sqrt{ss}}A_1 \ ,\nonumber\\
\text{A}_6 & = & \, \epsilon^2 \, \frac{ss-2m_q^2}{ss+2m_q^2}\left(6\epsilon m_q^2 E_4 - 4ss^2 E_5+m_q^2(3ss+2m_q^2)E_6\right)+\frac{14ss\,  m_q^2-13ss^2+8m_q^4}{3m_q^2(ss+2m_q^2)}A_1\nonumber\\
& & +\frac{6m_q^2}{ss+2m_q^2}F_{12} \ ,\nonumber
\label{be12}
\eea

\bea
\text{A}_7 & = & \, \epsilon^3 \,  (ss-2m_q^2)\,  E_7 \ , \nonumber\\
\text{A}_8 & = & \, \epsilon^3 \,  (ss-2m_q^2)\,  E_8 \ , \nonumber\\
\text{A}_9 & = & \, \epsilon^2 \, \sqrt{ ss}\sqrt{ss - 2 m_q^2}(ss-2m_q^2)E_9 + 4\frac{(2ss-5m_q^2)\sqrt{ss - 2 m_q^2}}{3m_q^2\sqrt{ss}}A_1\ ,\nonumber\\
\text{A}_{10} & = & \, \epsilon^3 \, (1-2\epsilon)(ss-m_q^2)E_{10} \ , \nonumber\\
\text{A}_{11} & = & \, \epsilon^4 \, (ss-2m_q^2)\,  E_{11} \ , \nonumber\\
\text{A}_{12} & = & \, \epsilon^3 \, \left(m_q^2(ss-2m_q^2)E_{12}-4\epsilon\, m_q^2 E_{11}\right)+\frac{2}{3}(\frac{ss}{m_q^2}-10)A_1+\frac{m_q^2(2F_{24}-4A_{7})}{ss-2m_q^2} \ , \nonumber\\
\text{A}_{13} & = & \, \epsilon^4 \, (ss-2m_q^2)\,  E_{13} \ , \nonumber\\
\text{A}_{14} & = & \, \epsilon^3 \, \left(m_q^2(ss-2m_q^2)E_{14}+4\epsilon\, m_q^2 E_{13}\right)-\frac{2}{3}(\frac{ss}{m_q^2}-10)A_1\nonumber\\
& & -\frac{m_q^2(2F_{27}-4A_{8})}{ss-2m_q^2} +\frac{2\sqrt{ss}}{\sqrt{ss-2m_q^2}}A_{9} \ , \nonumber\\
\text{A}_{15} & = & \, \epsilon^4 \, (ss-2m_q^2)\,  E_{15} \ , \nonumber\\
\text{A}_{16} & = & \, \epsilon^4 \, (ss-2m_q^2)\,  E_{16} \ , \nonumber\\
\text{A}_{17} & = & \, \epsilon^3 \, \sqrt{ ss}\sqrt{ss - 2 m_q^2}(ss-2m_q^2)\,  E_{17} \ , \nonumber\\
\text{A}_{18} & = & \, \epsilon^2 m_q^2(ss-2m_q^2)E_{18}-4\epsilon^4 m_q^2 E_{16}-\epsilon^3(3ss^2-8ss m_q^2+4m_q^4)E_{17}\nonumber\\
& & -\frac{\sqrt{ss-2m_q^2}}{\sqrt{ss}}(F_7+F_8+2F_9)+\frac{2m_q^2}{ss-2m_q^2}(F_{20}-2A_4)\nonumber\\
& & -\frac{ss+2m_q^2}{\sqrt{ss}\sqrt{ss-2m_q^2}}A_5+\frac{2(ss-10m_q^2)}{3ss}A_1\ , \nonumber\\
\text{A}_{19} & = & \, \epsilon^3 \, \sqrt{ss+2m_q^2}\sqrt{ss-2m_q^2}(ss-2m_q^2)\,  E_{19} \ , \nonumber\\
\text{A}_{20} & = & \, \epsilon^3 \, \sqrt{ss+2m_q^2}\sqrt{ss-2m_q^2}(ss-2m_q^2)\,  E_{20} \ , \nonumber\\
\text{A}_{21} & = & \, \epsilon^4 \, (ss-2m_q^2)\,  E_{21} \ , \nonumber\\
\text{A}_{22} & = & \, \epsilon^4 \, (ss-2m_q^2)\,  E_{22} \ , \nonumber\\
\text{A}_{23} & = & \, \epsilon^3 \, \sqrt{ss}\sqrt{ss-2m_q^2}(ss-2m_q^2)\,  E_{23} \ , \nonumber\\
\text{A}_{24} & = & \, \epsilon^3 \,(ss\, E_{24}-\frac{ss}{2}\, (ss-2m_q^2)E_{23}-2\epsilon\, m_q^2\, E_{22})-\frac{m_q^2}{ss-2m_q^2}\, A_{7}\ , \nonumber\\
\text{A}_{25} & = & \, \epsilon^4 \, (ss-2m_q^2)\,  E_{25} \ , \nonumber\\
\text{A}_{26} & = & \, \epsilon^3 \, \sqrt{ss+6m_q^2}\sqrt{ss-2m_q^2}(ss-2m_q^2)\,  E_{26} \ , \nonumber\\
\text{A}_{27} & = & \, \epsilon^3 \, (m_q^2 E_{27}-\frac{ss^2-4m_q^4}{2}E_{28})+\frac{m_q^2}{ss-2m_q^2}\left(F_{32}-F_{30}-2A_{7}\right)\, \ , \nonumber\\
\text{A}_{28} & = & \, \epsilon^3 \, \sqrt{ss+2m_q^2}\sqrt{ss-2m_q^2}(ss-2m_q^2)\,  E_{28} \ , \nonumber
\eea
\bea
\text{A}_{29} & = & \, \epsilon^4 \, (ss-2m_q^2)^2\,  E_{29} \ , \nonumber\\
\text{A}_{30} & = & \, \epsilon^4 \, (ss-2m_q^2)^2\,  E_{30} \ , \nonumber\\
\text{A}_{31} & = & \, \epsilon^4 \, (ss-2m_q^2)^2\,  E_{31} \ , \nonumber\\
\text{A}_{32} & = & \, \epsilon^4 \, (ss-2m_q^2)^2\,  E_{32} \ , \nonumber\\
\text{A}_{33} & = & \, \epsilon^4 \, \sqrt{ss+2m_q^2}\sqrt{ss-2m_q^2}(ss-2m_q^2)\,  E_{33} \ , \nonumber\\
\text{A}_{34} & = & \, \epsilon^4 \, (ss-2m_q^2)(E_{34}-(ss-2m_q^2) E_{33}) \ , \nonumber\\
\text{A}_{35} & = & \, \epsilon^4 \, \sqrt{ss+2m_q^2}\sqrt{ss-2m_q^2}(ss-2m_q^2)\,  E_{35} \ , \nonumber\\
\text{A}_{36} & = & \, \epsilon^4 \, (ss-2m_q^2)(E_{36}-(ss-2m_q^2) E_{35}) \ , \nonumber\\
\text{A}_{37} & = & \, \epsilon^4 \, \sqrt{ss+2m_q^2}\sqrt{ss-2m_q^2}(ss-2m_q^2)\,  E_{37} \ , \nonumber\\
\text{A}_{38} & = & \, \epsilon^4 \, (ss-2m_q^2)(E_{38}+(ss+2m_q^2) E_{37}) \ , \nonumber\\
\text{A}_{39} & = & \, \epsilon^4 \, \sqrt{ss+2m_q^2}\sqrt{ss-2m_q^2}(ss-2m_q^2)^2\,  E_{39} \ .
\label{base}
\eea

With the basis chosen above, the differential equations for $(\text{A}_3\ldots \text{A}_{39})$ then turn to the canonical form, except for $\text{A}_5$ and $\text{A}_9$. The differential equations for $\text{A}_5$ and $\text{A}_9$ with respect to $x$ write as:
\bea
\frac{\text{d} \text{A}_5}{\text{d} x} &=&\epsilon\frac{33\text{A}_4+6\text{A}_5-6\text{A}_6-4F_8+21F_{12}}{4x}+\epsilon\frac{\text{A}_5+2F_8}{x-1}\nonumber\\
& &+\epsilon\frac{9\text{A}_4-2\text{A}_5-2\text{A}_6+9F_{12}}{x-3}+\epsilon\frac{9\text{A}_4+2\text{A}_5-2\text{A}_6+9F_{12}}{x-\frac{1}{3}}\nonumber\\
&&+\epsilon\frac{1}{6}\left(\frac{1}{x^2}+\frac{28}{x}-\frac{40}{(x-1)^2} + \frac{80}{x-3} - \frac{80}{x - \frac{1}{3}}+1\right)A_1\nonumber\\
&&-\frac{4}{3}\left(\frac{5}{(x-1)^2}+\frac{1}{x}\right)\text{A}_1\ ,\nonumber\\
\frac{\text{d} \text{A}_9}{\text{d} x} &=&\epsilon\frac{6\text{A}_8+3\text{A}_9+2F_7}{x}-\epsilon\frac{4\text{A}_9+F_7}{x-1}-\epsilon\frac{2\text{A}_9}{x+1}\nonumber\\
& & +\epsilon\frac{1}{3}\left(\frac{4}{x^2}+\frac{160}{(x-1)^2}+\frac{52}{x}+4\right)\text{A}_1\nonumber\\
& & -\frac{8}{3}\left(\frac{5}{(x-1)^2}+\frac{1}{x}\right)\text{A}_1\ .
\eea
Notice that the above two equations are not in canonical form, and they both have the $\epsilon$ free $\text{A}_1$ terms, by a factor of 2 difference. Those terms without $\text{A}_1$ can be expressed in d-log form.
By using the method described in above, different from casting all terms into canonical form via (non-algebraic) basis change in Ref. \cite{adams:2018}, the obtained differential equations are greatly simplified and are suitable for solving recursively. Taking the known result on $\text{A}_1$ \cite{Remiddi:2016gno} as an input, the differential equations for $(\text{A}_3\ldots \text{A}_{39})$ can be integrated straightforwardly order by order in $\epsilon$. The corresponding lengthy expressions is given as an auxiliary file in {\bf arXiv} version of this paper.

After determining the bases, to fix the boundary conditions is necessary for solving the differential equations. Here, we apply the regularity conditions as in Ref. \cite{Gehrmann:1999as} to assist the determination of boundary conditions. Noticing that the integrals ($E_3,$ $E_4,$ $E_5,$ $E_7,$ $E_8,$ $E_9,$ $E_{11},$ $E_{13},$ $E_{15},$ $E_{16},$ $E_{17},$ $E_{19}\ldots E_{23},$ $E_{25},$ $E_{26},$ $E_{28}\ldots E_{39}$) are regular at $ss=2m_q^2$ and multiplying the normalization factor $(ss-2m_q^2)$ to $\text{A}_i$, one may find that the corresponding bases $\text{A}_i$ turn to be zero at $ss=2m_q^2$. The boundary condition for $\text{A}_6$ at $ss=2m_q^2$ can be fixed in a similar way, that is
\bea
\text{A}_6\mid_{ss=2m_q^2}=\frac{3}{2}F_{12}-\frac{4}{3}\text{A}_1\mid_{ss=2m_q^2}=\frac{1}{2}F_{12}\ .
\eea
Here, the integral $F_{12}$ is known, and the boundary condition for $\text{A}_1$ may be determined from its definition in (\ref{be12}), i.e. $\text{A}_1\mid_{ss=2m_q^2}=\frac{3}{4}F_{12}$. The integral $E_{10}$ is regular at $ss=m_q^2$ with the normalization factor $(ss-m_q^2)$. Multiplied by this normalization factor, we then find $\text{A}_{10}=0$ at $ss=m_q^2$. Since the integrals $(E_{12},E_{14},E_{18},E_{27})$ are also regular at $ss=2m_q^2$,  the boundaries of corresponding bases $\text{A}_i$ can be determined by differential equations. For instance, the differential equation for $\text{A}_{12}$ reads
\bea
\frac{\text{d} \text{A}_{12}}{\text{d} y} = 2\epsilon\frac{6F_{24}-12\text{A}_7 + 6\text{A}_{11}-3\text{A}_{12}-16\text{A}_1}{3 y}+\ldots \ ,
\label{A12}
\eea
where ellipses stand for less singular terms at $y=0$, i.e. $ss=2m_q^2$. Since all integrals in (\ref{A12}) have finite limits at $y\rightarrow 0$, the following relation between different integrals exists:
\bea
\lim_{y\rightarrow 0} (6F_{24}-12\text{A}_7 + 6\text{A}_{11}-3\text{A}_{12}-16\text{A}_1)=0.
\eea
Because $(F_{24},\text{A}_7,\text{A}_{11})$ are zero at $y=0$ ($ss=2m_q^2$), we then have
\bea
\text{A}_{12}\mid_{y=0}=-\frac{16}{3}\text{A}_{1}\mid_{y=0}\ .
\eea
Similarly, from those boundaries for integrals $E_{14}$, $E_{18}$, $E_{24}$ and $E_{27}$, one can fix all boundary conditions for bases $(\text{A}_1\ldots \text{A}_{39})$, of which the none-zero ones up to weight-4 write as:
\bea
\text{A}_1\mid_{ss=2m_q^2} &=& \epsilon^2\frac{\pi^2}{16}+\epsilon^3\frac{3}{16}(7\zeta(3)-2\pi^2\ln(2))+\epsilon^4(9\text{Li}_4(\frac{1}{2})-\frac{31\pi^4}{480}+\frac{3}{4}\pi^2\ln^2(2)\nonumber\\
& & +\frac{3}{8}\ln^4(2))+ {\cal O}(\epsilon^{5}),\nonumber\\
\text{A}_2\mid_{ss=2m_q^2} &=& \frac{16}{3}\text{A}_1\mid_{ss=2m_q^2},\nonumber\\
\text{A}_6\mid_{ss=2m_q^2} &=& \frac{2}{3}\text{A}_1\mid_{ss=2m_q^2},\nonumber\\
\text{A}_{12}\mid_{ss=2m_q^2} &=& -\frac{16}{3}\text{A}_1\mid_{ss=2m_q^2},\nonumber\\
\text{A}_{14}\mid_{ss=2m_q^2} &=& \frac{8}{3}\text{A}_1\mid_{ss=2m_q^2},\nonumber\\
\text{A}_{18}\mid_{ss=2m_q^2} &=& -\frac{4}{3}\text{A}_1\mid_{ss=2m_q^2},\nonumber\\
\text{A}_{24}\mid_{ss=2m_q^2} &=& 8\epsilon^3\pi^2\ln(2)+\epsilon^4(\frac{59\pi^4}{15}-192\text{Li}_4(\frac{1}{2})-8\ln^2(2)(\pi^2+\ln^2(2)))+ {\cal O}(\epsilon^{5}),\nonumber\\
\text{A}_{27}\mid_{ss=2m_q^2} &=& \epsilon^3(\frac{3\zeta(3)}{2}+\pi^2\ln(2))+\epsilon^4(-24\text{Li}_4(\frac{1}{2})+\frac{19\pi^4}{30}-\ln^4(2))+ {\cal O}(\epsilon^{5}).
\eea

\subsection{Sector II : non-planar two-loop three-point integrals}

\begin{figure}[h]
\begin{center}
\includegraphics[scale=0.47]{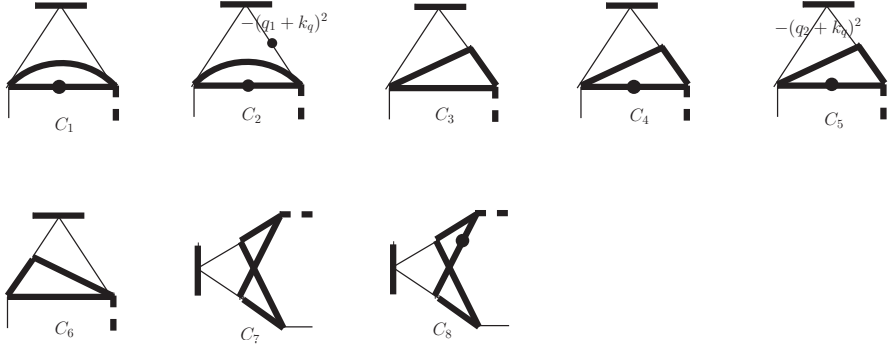}
\caption{The set of 8 master integrals that appear in sector II. Integrals $(C_1\ldots C_6)$ can be cast into canonical form, while integrals $(C_7,C_8)$ involve elliptic functions. The thin line denotes massless propagators and on-shell massless external particles; the thick line presents massive propagators and on-shell massive external particles; the dash line indicates off-shell external particles with momentum squared equal to $2 ss$. The internal lines with a dot mean the power of the propagators being raised to 2.}
\label{midiag2}
\end{center}
\end{figure}

In this subsection we consider the non-planar two-loop three-points integrals that appear in the massive light-by-light Feynman diagrams. There are eight master integrals, as shown in Fig. \ref{midiag2}, with the corresponding bases $B_i$ as
\bea
B_1 & = & \, \epsilon^3 \, (ss-2m_q^2)\,  C_{1} \ , \nonumber\\
B_2 & = & \, \epsilon^2 \, \frac{-\sqrt{(ss-2m_q^2)(ss-4m_q^2)}}{4m_q^2}(2(ss-2m_q^2)C_{1}-4m_q^2 C_{2}+F_{6}/\epsilon^2) \ , \nonumber\\
B_3 & = & \, \epsilon^4 \, (ss-2m_q^2)\,  C_{3} \ , \nonumber\\
B_4 & = & \, \epsilon^3 \, \sqrt{ss}\sqrt{ss-2m_q^2}(ss-2m_q^2)\,  C_{4} \ , \nonumber\\
B_5 & = & \, \epsilon^3 \, ss(C_{5}-\frac{(ss-2m_q^2)}{2}C_{4}-\epsilon\,  C_{3}-\frac{C_1}{2}) \ , \nonumber\\
B_6 & = & \, \epsilon^4 \, (ss-2m_q^2)\,  C_{6} \ , \nonumber\\
B_7 & = & \, \epsilon^4 \, (ss-2m_q^2)^2\,  C_{7} \ , \nonumber\\
B_8 & = & \, \epsilon^4 \, \frac{ss^2-4ss m_q^2+20m_q^4}{ss-2m_q^2}\, m_q^4  C_{8} \ .
\label{baseb}
\eea
Note, here the integrals ($C_1\ldots C_6$) were first calculated in Ref. \cite{Bonciani:2016qxi}, and the left two non-planar two-loop integrals $(B_7,B_8)$ cannot be cast into the canonical form via algebraic change of basis. A similar topology of Feynman diagram as that of ($C_7,C_8$), but with different kinematics and outgoing momentum squared, was handled in Ref. \cite{vonManteuffel:2017hms}.

In order to get expressions for $B_7$ and $B_8$ we first derive two coupled first-order differential equations with the evolution of variable $ss$, and then transform them to a second-order differential equation for $B_7$. That is:
\bea
& &\frac{\text{d}^2 B_7 }{d ss^2}-\frac{ss^2-4ss m_q^2-12m_q^4}{(ss-2m_q^2)(ss^2-4ss m_q^2+20m_q^4)} \frac{\text{d} B_7}{\text{d} ss}\nonumber\\
&&-\frac{16 m_q^4}{(ss-2m_q^2)^2(ss^2-4ss m_q^2+20m_q^4)}B_7=N(\epsilon,ss,m_q^2)\ ,
\eea
with $N(\epsilon,ss,m_q^2)$ denoting the non-homogeneous term. Here, the tough issue is how to determinate the homogeneous solution. To this aim, we make a variable transformation of $ss$ to $v=\frac{-i (ss-2 m_q^2)}{4 m_q^2}$, then the homogenous part of the differential equation turns to
\bea
\frac{\text{d}^2 B_7 }{d v^2}-\frac{1+v^2}{v(1-v^2)} \frac{\text{d} B_7}{\text{d} v}+\frac{1}{v^2(1-v^2)} B_7=0\ .
\label{simpeq}
\eea

The solutions of equation (\ref{simpeq}) can be readily obtained. The two homogeneous solutions $(y_1(v),y_2(v))$ read
\bea
y_1(v)=v K(v^2),~~~~y_2(v)=v K(1-v^2),
\eea
with $K(x)$ being the first kind complete elliptic integral. Note that the recently development on maximal-cut \cite{Harley:2017qut,Primo:2016ebd,Bosma:2017ens} is also applicable to the determination of the homogeneous solution. The Wronskian of the homogeneous solution reads
\bea
w(v)=y_2(v)\frac{\text{d} y_1(v)}{\text{d} v}-y_1(v)\frac{\text{d} y_2(v)}{\text{d} v}=\frac{v \pi}{2(1-v^2)}\ .
\eea
With the homogeneous solutions and Wronskian, a particular solution can be obtained by means of the constant variation. The general solution is then
\bea
B^i_7=c_1 y_1(v) + c_2 y_2(v)- y_1(v)\int_{0}^{v}\frac{N^i(\alpha)}{w(\alpha)}y_2(\alpha)\text{d}\alpha + y_2(v)\int_{0}^{v}\frac{N^i(\alpha)}{w(\alpha)}y_1(\alpha)\text{d}\alpha\ ,
\eea
where $i$ refers to the order of $\epsilon$ in $B_7$.

Since the integral $C_7$ has no singularity at $ss=2m_q^2$, and the normalization for $C_7$ in $B_7$ is $(ss-2m_q^2)^2$, we know
\bea
B_7\mid_{(v=0)}=0,~~~~\frac{\text{d} B_7}{\text{d} v}\mid_{(v=0)}=0\ .
\eea
Hence, the constants $c_1$ and $c_2$ can be fixed to
\bea
c_1=c_2=0\ .
\eea
Once $B_7$ is obtained, we can then determine the $B_8$ from the first order differential equation with respect to $B_7$ straightforwardly.

Before calculating the differential equations for integrals in this sector, still the corresponding boundary conditions should be fixed. Since the integrals $(B_1, B_2, B_3, B_4, B_6)$ are regular at $ss=2m_q^2$, by multiplying
their normalization factor $(ss-2m_q^2)$  to $B_i$, the corresponding bases $B_i$ then turn out to be zero at $ss=2m_q^2$. Considering that the master integrals in basis $B_5$ are regular as $ss=0$ and have a common normalization factor $ss$, we readily know $B_5=0$ when $ss=0$. With these discussions, all necessary boundary conditions to fix the solutions of differential equations are ready.

\subsection{Analytic continuation and discussions}

With the analytical results obtained in above, the next necessary step is to determinate the analytic continuation of the master integrals, which is similar to the procedure in our previous work \cite{Chen:2017xqd}.  The correct analytic continuation can be achieved by the replacement of $ss \rightarrow ss + i 0$ at fixed $m_q^2$, which corresponds to $x\rightarrow x + i 0$, $y\rightarrow y + i 0$ and $z\rightarrow z + i 0$.

The canonical bases in (\ref{base}) contain 4 independent square roots
\be
(\sqrt{ss},\sqrt{ss-2m_q^2},\sqrt{ss+2m_q^2},\sqrt{ss+6m_q^2})\ ,
\ee
which cannot be simultaneously rationalized via one variable change. This means it is not possible to
integrate the differential equations directly in terms of Gongcharov polylogarithms. It is worth mentioning that Refs. \cite{Bonciani:2016qxi,vonManteuffel:2017myy} proposed some novel ways to express the results of canonical bases for non-elliptic sectors in terms of multiple polylogarithms, without considering the existence of rational parametrization of the alphabet. However the results tend to be rather lengthy when expressed in multiple polylogarithms. In order to calculate the integrals numerically in a faster and convenient way, we construct a one-fold integral representation for the integrals that can be cast into the canonical form by means of what proposed in Ref. \cite{Bonciani:2016qxi}. For integrals in elliptic sectors we need the two-fold integral representation to express the results up to weight four. The one fold and two fold integral representations we adopted are suitable for fast and precise numerical evaluation with Mathematica program on a single core computer.

The analytic calculation in this work is performed by our own developed Mathematica code, and in order to guarantee the correctness of our results, we ask all analytical expressions for master integrals experiencing at least one independent examination. We check all results in contrast to those obtained via numerical programs {\bf Fiesta} \cite{Smirnov:2013eza,Smirnov:2015mct} and {\bf SecDec} \cite{Borowka:2012yc,Borowka:2015mxa}. We have achieved an excellent agreement in analytical and numerical approaches with kinematics in both Euclidean and Minkowski regions.

\section{Conclusions and outlooks}

The integrals involving elliptic functions in the NNLO QCD corrections to heavy quarkonium exclusive production and decays are calculated, which turns out to be a tough issue. Those integrals are classified into two sectors, one with integrals containing sub-topologies related to the two-loop massive sunrise integrals and another with two massive two-loop non-planar three-points integrals. We find the simple example studied in Ref. \cite{Remiddi:2016gno} is in fact applicable to more general cases, that is, the expressions for two master integrals composed of two-loop massive sunrise integrals are still suitable for our case. In order to compute the first sector Feynman integrals under consideration we exploit the result for the two-loop massive sunrise integrals in Ref. \cite{Remiddi:2016gno}. We find a suitable linear combination of Feynman integrals such that only one of the master integrals about the solutions of two-loop massive sunrise integrals is required. By properly choosing canonical basis, we transform the differential equations into a simple and compact form that can be solved recursively. For another elliptic sector, the key point is to solve the homogeneous equation, with that inhomogeneous solutions can be obtained by means of constant variation.

Together with those 86 integrals calculated in our previous work \cite{Chen:2017xqd}, all master integrals appearing in the calculation of NNLO QCD correction to CP-even heavy quarkonium exclusive production and decays, such as $\gamma^*\gamma\rightarrow Q\bar{Q}$ and $e^+e^-\rightarrow \gamma+ Q\bar{Q}$ \cite{Chen:2017pyi}, are ready. The master integrals take the form of mutilple polylogarithms, iterative integrals over complete elliptic integrals and multiple polylogarithms. It is noteworthy that the integrals calculated in this work may also appear in the calculation of NNLO corrections in other processes, such as the exclusive decay of Higgs or $Z^0$ boson to quarkonium plus a photon and the inclusive hadronic production or decay of $\eta_c/\eta_b$, which are also phenomenologically meaningful. Moreover, we tend to believe that the calculation procedure and results in this work might be helpful to the mater integrals calculation of processes beyond the scope of heavy quarkonium physics, for instance the NNLO corrections to top quark pairs hadronic production, and NNLO corrections to heavy quark pair production plus a jet in electron-positron collision.

Note, only simple results are given in the appendix, however the full but lengthy results will be provided upon request.

\acknowledgments

This work was supported in part by the Ministry of Science and Technology of the People's Republic of China(2015CB856703); by the Strategic Priority Research Program of the Chinese Academy of Sciences, Grant No.XDB23030100; and by the National Natural Science Foundation of China(NSFC) under the grants 11375200 and 11635009. We are grateful to the anonymous referee for valuable comments and suggestions.

\appendix
\section{The definition for integrals}

The integral $\text{A}_1$ is defined as
\be
\text{A}_1 = \int {\mathcal D}^Dq_1 \, {\mathcal D}^Dq_2 \, \frac{1}{-q_1^2+m_q^2} \, \frac{1}{-q_2^2+m_q^2} \, \frac{1}{-(q_1+q_2+k_1-k_q)^2+m_q^2}\  ,
\ee
where the measure of the integration is
\be
{\mathcal D}^Dq_i = \frac{1}{\pi^{D/2}\Gamma(1+\epsilon)}\left(\frac{m_q^2}{\mu^2}\right)^\epsilon  d^Dq_i \ .
\ee
For master integrals without numerators, their definition can be read off from Fig. \ref{midiag1} and Fig. \ref{midiag2}, with the normalization defined in above. For master integrals with numerators, we can define a series of propagators as
\begin{align}
P_{1} & =m_{q}^{2}-q_{1}^{2},\hspace{3.1cm}P_{2}=m_{q}^{2}-q_{2}^{2},\nonumber \\
P_{3} & =-(q_{1}+q_{2})^{2},\hspace{2.52cm}P_{4}=m_{q}^{2}-(q_{1}+k_{1})^{2},\nonumber \\
P_{5} & =m_{q}^{2}-(q_{2}+k_{2})^{2},\hspace{1.82cm}P_{6}=m_{q}^{2}-(q_{1}+k_{2})^{2},\nonumber \\
P_{7} & =-(q_2-k_q)^2,\hspace{2.57cm}P_{8}=-(q_{2}+k_{q}-k_2)^{2},\nonumber \\
P_{9} & =-(q_{2}+k_2-k_{q})^{2},\hspace{1.63cm}P_{10}=-(q_1+q_{2}+k_{q})^{2}+m_q^2,\nonumber \\
P_{11} & =-(q_{1}+k_{q})^{2},\hspace{2.5cm}P_{12}=-(q_{2}+k_{q})^2,\nonumber\\
P_{13} &= -(q_1+k_1+k_q)^2, \hspace{1.7cm}P_{14}=-(q_1-k_q)^2,\nonumber\\
P_{15} &=-(q_2+k_1)^2+m_q^2\ .
\end{align}
Then, the master integrals with numerators can be expressed as
\bea
M_{24} & = & \int {\mathcal D}^Dq_1 \, {\mathcal D}^Dq_2 \frac{P_{11}}{P_1 P_2 P_4^2 P_7 P_{10}}, ~~~~~~~~
M_{34}  =  \int {\mathcal D}^Dq_1 \, {\mathcal D}^Dq_2 \frac{P_{7}}{P_1 P_2 P_3 P_4 P_9 P_{10}}, \nonumber\\
M_{36} & = & \int {\mathcal D}^Dq_1 \, {\mathcal D}^Dq_2 \frac{P_{12}}{P_1 P_2 P_3 P_6 P_8 P_{10}}, ~~~~
M_{38}  = \int {\mathcal D}^Dq_1 \, {\mathcal D}^Dq_2 \frac{P_{12}}{P_1 P_2 P_3 P_4 P_5 P_{10}},\nonumber\\
C_{2} & = &  \int {\mathcal D}^Dq_1 \, {\mathcal D}^Dq_2 \frac{P_{11}}{ P_2 P_9 P_{10} P_{13}},~~~~~~~~~
C_{5}  =   \int {\mathcal D}^Dq_1 \, {\mathcal D}^Dq_2 \frac{P_{12}}{ P_2 P_{10} P_{11} P_{14} P_{15} }\ .
\eea

\section{The typical analytical results}

The typical analytic results of the 39 canonical bases $A_i$, in terms of GPLs and iterative integrals over complete elliptic integrals, are:

\allowdisplaybreaks{
\begin{flalign}
 & \text{A}_{1}=\epsilon^2 \int_9^{\infty}\frac{\text{d} t}{t-y-1-i \varepsilon} I_1(t)+{\cal O}(\epsilon^{3})\ , \nonumber\\
 & \text{A}_{2}=\epsilon^2 \left( \frac{1}{\sqrt{3}}\text{Cl}(\frac{\pi}{3})+(y+1)(\frac{5}{6}+\sqrt{3}\text{Cl}(\frac{\pi}{3}))+(y+1)^2 \int_9^{\infty}\frac{\text{d} t}{t^2(t-y-1-i \varepsilon)} I_2(t) \right)\nonumber\\
 &+{\cal O}(\epsilon^{3})\ ,\nonumber\\
 & \text{A}_{3}=\epsilon^3 \int_9^{\infty} -\frac{4}{3}G(t-1,y)I_1(t) \text{d} t+{\cal O}(\epsilon^{4})\ ,\nonumber\\
 & \text{A}_{4}={\cal O}(\epsilon^{3})\ ,\nonumber\\
 & \text{A}_{5}=\epsilon^2\big(G(1,0,x)-\frac{1}{2}G(0,0,x)-\frac{\pi^2}{6}-\ln(2)\pi i+\frac{10(x+1)}{3(x-1)}\text{E}_c \nonumber\\
 &+\int_9^{\infty} \frac{2(2t-3)\ln(-\frac{x+t+\sqrt{t^2-1}}{1+(t+\sqrt{t^2-1})x})}{3(t+1)\sqrt{t^2-1}}I_1(t)  \text{d} t\big)+{\cal O}(\epsilon^{3})\ ,\nonumber\\
 & \text{A}_{6}=\epsilon^{2}(\frac{1}{2}G(0,0,x)+\frac{7 \pi^2}{24})+{\cal O}(\epsilon^{3})\ ,\nonumber\\
 & \text{A}_{7}=\epsilon^3 \int_9^{\infty} -\frac{8}{3}G(t-1,y)I_1(t) \text{d} t+{\cal O}(\epsilon^{4})\ ,\nonumber\\
 & \text{A}_{8}={\cal O}(\epsilon^{3})\ ,\nonumber\\
 & \text{A}_{9}=\epsilon^2\big(2G(1,0,x)-G(0,0,x)-\frac{\pi^2}{3}-2\ln(2)\pi i+\frac{20(x+1)}{3(x-1)}\text{E}_c\nonumber\\
 &+\int_9^{\infty} \frac{4(2t-3)\ln(-\frac{x+t+\sqrt{t^2-1}}{1+(t+\sqrt{t^2-1})x})}{3(t+1)\sqrt{t^2-1}}I_1(t)  \text{d} t\big)+{\cal O}(\epsilon^{3})\ ,\nonumber\\
 & \text{A}_{10}=\epsilon^3\big(2G(-1,0,0,y)-G(0,-1,0,y)-i \pi(G(0,-1,y)-2G(-1,0,y)-G(-1,y)\pi^2 \nonumber\\
 & -\frac{3}{2}\zeta(3)+\pi^2\ln(2)) + \int_9^{\infty} -\frac{2(t-9)G(t-1,y)}{3(t-1)}I_1(t)  \text{d} t\big)+{\cal O}(\epsilon^{4})\ , \nonumber\\
 & \text{A}_{11}={\cal O}(\epsilon^{4})\ ,\nonumber\\
 & \text{A}_{12}=\epsilon^2(-G(-1,0,y)+i\pi G(-1,y)-\frac{\pi^2}{3})+\epsilon^3(4G(-1,0,0,y)+2G(0,-1,0,y)\nonumber\\
 & -3G(-1,-1,0,y)-2G(1,0,0,y)+i\pi(3G(-1,-1,y)-4G(-1,0,y)+2G(1,0,y)\nonumber\\
 &-2G(0,-1,y))+\frac{2}{3}\pi^2G(0,y)-\frac{3}{2}\pi^2G(-1,y)-7\zeta(3)+2\pi^2\ln(2)\nonumber\\
 & +\int_9^{\infty}\frac{4((t+7)G(t-1,y)-8G(0,y))}{3(t-1)}I_1(t)  \text{d} t)+ {\cal O}(\epsilon^{4})\ ,\nonumber\\
 & \text{A}_{13}={\cal O}(\epsilon^{4})\ ,\nonumber\\
 & \text{A}_{14}=\epsilon^2\big(4G(i,-1,x)+4G(-i,-1,x)-2G(i,0,x)-2G(-i,0,x)-4G(0,-1,x)\nonumber\\
 &-2G(0,0,x) +2\ln(2)(G(0,x)-G(-i,x)-G(i,x))-\pi^2+\ln^2(2)\big)/2+{\cal O}(\epsilon^{3})\ ,\nonumber\\
 & \text{A}_{15}=\epsilon^4\big[G(0,0,0,1,z)-2G(1,0,0,1,z)+\zeta(3)G(0,z)-2\zeta(3)G(1,z)-\frac{\pi^4}{60}\nonumber\\
 & +\int_9^{\infty}\frac{2}{3}(4G(1,t,z)-3G(0,t,z)+G(t,1)(3G(0,z)-4G(1,z))\nonumber\\
 &-4\text{Li}_{2}(\frac{1}{1-t})-3\text{Li}_{2}(\frac{1}{t}))I_1(t)  \text{d} t)\big]+ {\cal O}(\epsilon^{5})\ ,\nonumber\\
 & \text{A}_{16}=  {\cal O}(\epsilon^{4})\ ,\nonumber\\
 & \text{A}_{17}=  {\cal O}(\epsilon^{3})\ ,\nonumber\\
 & \text{A}_{18}= \epsilon^2\big(2G(0,-1,x)-2G(-i,-1,x)-2G(i,-1,x)+G(i,0,x)+G(-i,0,x)\nonumber\\
 &+\ln(2)(G(i,x)+G(-i,x)-G(0,x))+\frac{\pi^2}{12}-\frac{\ln^2(2)}{2}\big)+ {\cal O}(\epsilon^{3}),\nonumber\\
 & \text{A}_{19}=  {\cal O}(\epsilon^{3})\ ,\nonumber\\
 & \text{A}_{20}=  {\cal O}(\epsilon^{3})\ ,\nonumber\\
 & \text{A}_{21}=  {\cal O}(\epsilon^{4})\ ,\nonumber\\
 & \text{A}_{22}=  {\cal O}(\epsilon^{4})\ ,\nonumber\\
 & \text{A}_{23}=  {\cal O}(\epsilon^{3})\ ,\nonumber\\
 & \text{A}_{24}=  {\cal O}(\epsilon^{3})\ ,\nonumber\\
 & \text{A}_{25}=  {\cal O}(\epsilon^{4})\ ,\nonumber\\
 & \text{A}_{26}=  {\cal O}(\epsilon^{3})\ ,\nonumber\\
 & \text{A}_{27}=  {\cal O}(\epsilon^{3})\ ,\nonumber\\
 & \text{A}_{28}=  {\cal O}(\epsilon^{3})\ ,\nonumber\\
 & \text{A}_{29}= \epsilon^4\big(2G(0,1,0,1,z)+2G(0,0,1,1,z)+4G(0,-1,0,1,z)-2G(0,0,0,1,z)\nonumber\\
 &-G(-1,0,-1,0,y)+i\pi G(-1,0,-1,y)+\frac{\pi^2}{3}(G(0,1,z)-G(0,-1,z))\nonumber\\
 &+\zeta(3)G(0,z)+\pi^2\ln(2)G(-1,y)-\frac{7\zeta(3)}{2}G(-1,y)+\frac{17\pi^4}{360} \nonumber\\
 &+\int_9^{\infty}\frac{-8(t-3)G(-1,t-1,y)}{3(t-1)}I_1(t)  \text{d} t\big)+{\cal O}(\epsilon^{5})\nonumber\\
 & \text{A}_{30}=  {\cal O}(\epsilon^{4})\ ,\nonumber\\
 & \text{A}_{31}=  {\cal O}(\epsilon^{4})\ ,\nonumber\\
 & \text{A}_{32}=  {\cal O}(\epsilon^{4})\ ,\nonumber\\
 & \text{A}_{33}=  {\cal O}(\epsilon^{4})\ ,\nonumber\\
 & \text{A}_{34}=  {\cal O}(\epsilon^{4})\ ,\nonumber\\
 & \text{A}_{35}=  {\cal O}(\epsilon^{4})\ ,\nonumber\\
 & \text{A}_{36}=  \epsilon^4\big(2G(0,0,1,1,z)+2G(0,1,0,1,z)+4G(0,-1,0,1,z)-4G(0,0,0,1,z)\nonumber\\
 & -2G(-1,-1,-1,0,y)-2G(-1,-1,0,0,y)+2 i \pi(G(-1,-1,-1,y)+G(-1,-1,0,y))\nonumber\\
 & +\frac{\pi^2}{3}(G(0,1,z)-G(0,-1,z)) + \frac{\pi^2}{3}(G(-1,0,y)+4G(-1,-1,y))-\frac{7\zeta(3)}{2}G(-1,y)\nonumber\\
 &+\pi^2\ln(2)G(-1,y)-\zeta(3)G(0,z)+\frac{\pi^4}{40}\nonumber\\
 &+\int_9^{\infty}\frac{-8((t-3)G(-1,t-1,y)+(t-1)G(0,t-1,y) + 2G(-1,0,y))}{3(t-1)}I_1(t) \text{d} t \big)+{\cal O}(\epsilon^{5})\ ,\nonumber\\
 & \text{A}_{37}=  {\cal O}(\epsilon^{4})\ ,\nonumber\\
 & \text{A}_{38}=  {\cal O}(\epsilon^{4})\ ,\nonumber\\
 & \text{A}_{39}=  {\cal O}(\epsilon^{4})\ .\nonumber\\
\end{flalign}
}
Here, the elliptic functions $I_1(t)$ and $J_1(t)$ were first defined in Ref. \cite{Remiddi:2016gno} and formulated as
\bea
I_1(t)&=&\frac{2}{\sqrt{(\sqrt{t}+3)(\sqrt{t}-1)^3}}K(\frac{(\sqrt{t}-3) (\sqrt{t}+1)^3}{(\sqrt{t}+3)(\sqrt{t}-1)^3})\ ,\nonumber\\
J_1(t)&=&\frac{2}{\sqrt{(\sqrt{t}+3) (\sqrt{t}-1)^3}}K(\frac{16\sqrt{u}}{(\sqrt{t}+3)(\sqrt{t}-1)^3})\ .
\eea
The constant $E_c$ is defined as
\bea
E_c=\int_9^{\infty} \frac{I_1(t)}{t+1} \text{d} t\ .
\eea

\vspace{3ex}
\bibliographystyle{JHEP}
\bibliography{references}

\end{document}